\documentclass{jaa}
\usepackage{graphicx}
\usepackage{url}
\usepackage{comment}
\usepackage{mathptmx}
\usepackage{setspace}
\usepackage{float}
\usepackage{amsmath,amssymb}
\usepackage{booktabs}
\usepackage{algorithm}
\usepackage{algpseudocode}
\usepackage{enumitem}
\usepackage{array}
\usepackage{natbib}
\usepackage{hyperref}
\hypersetup{colorlinks = true, urlcolor = blue, linkcolor = blue, citecolor = blue}
\usepackage[version=4]{mhchem}
\newcommand{\hi}{\mbox{H\,{\sc i}}}
\newcommand{\oi}{\mbox{O\,{\sc i}}}

\begin{document}\sloppy

%%paper title
%%For line breaks \\ can be used within title
\title{DewTwin-Coin: an onboard autonomous framework for lunar water-ice prospecting using Chandrayaan-3 LIBS and ChaSTE data}

%%author names are separated by comma (,)
%%use \and before the last author name
%%use a * along with the number separated by comma
%% for the  author for correspondence
%%\textsuperscript{number} is used for affiliation
%%\affilOne, \affilTwo etc., upto \affilTwentyfive is possible
%%Please note the first letter after \affil is capitalised in the command

\author{Soundarya R.\textsuperscript{1,*} and Debasish Mondal\textsuperscript{2}}
\affilOne{\textsuperscript{1}Department of Computer Science and Engineering, The Oxford College of Engineering, Bommanahalli, Bengaluru 560068, Karnataka, India\\}
\affilTwo{\textsuperscript{2}Department of Physics, Indian Institute of Science Education and Research (IISER) Tirupati, Yerpedu, Tirupati 517619, Andhra Pradesh, India\\}

%%escape two column mode for title, affiliation and abstract by giving \twocolumn command as shown
\twocolumn[{
\maketitle

%%include \corres to print the corresponding author Email id
\corres{soundaryaramachandra2003@gmail.com}

%%include \msinfo for manuscript information such as received, revised and accepted dates
%\msinfo{1 September 2026}{1 September 2026}

%%abstract
\begin{abstract}
It is important to locate water-ice in the lunar regolith to build futuristic, sustainable bases there. But communication delays from Earth, limited bandwidth, and power constraints demand an onboard, autonomous, time-efficient decision-making framework for locating volatiles such as water-ice in the lunar regolith. In this work, we present a framework, \textit{DewTwin-Coin}, that can locate potential water-ice sites in real time without drilling. Its decision-making principle is based on a \textit{find-S learning} algorithm that uses surface temperature and the Hydrogen-to-Oxygen intensity ratio at a location. The core idea is a physics-based rule -- water-ice is stable only if the temperature is below $110\,\mathrm{K}$ and the Hydrogen-to-Oxygen intensity ratio is between 1.7 and 2.3. We validate \textit{DewTwin-Coin} on Chandrayaan-3's 3165 laser-induced breakdown spectroscopy (LIBS) elemental data with 387 Chandra's surface thermophysical experiment (ChaSTE) data. On a 16 GB RAM system, it classified all 3165 locations within 615 seconds, yielding no strong signatures of water-ice, which matches the Chandrayaan-3 in-situ analysis. This framework is very important for future lunar and other planetary missions in which an onboard rover can autonomously determine feasible drilling locations for water-ice without issuing false drilling commands in warm terrains. Thus, the mission's onboard setup can stretch its technical limits in both power and memory capacity.
\end{abstract}

%%insert keywords separated by 3 hyphens using \keywords{words}
}]
%%close the twocolumn escape here

%%include \doinum{number}for the DOI number in the header
%%include \volnum{number} for the volume number in the header
%%include \year{yyyy} for  year of publication in the header
%%include \pgrange{num--num} page range of article in the header
%%include \artcitid{num} for the article citation id
%%include \lp to print last page of the article
%%include \setcounter{page}{pagenum} for the exact starting page of the article

\doinum{12.3456/s78910-011-012-3}
\artcitid{\#\#\#\#}
\volnum{000}
\year{0000}
\pgrange{1--}
\setcounter{page}{1}
\lp{1}

\section{Introduction}
\subsection{Importance of lunar missions}
Water is the most valuable resource for sustaining life on Earth, yet it remains obscured elsewhere in the universe. Since the first lunar mission, \texttt{Apollo 11}, launched on July 20, 1969 \citep{Floca2019}, astronomers have been searching for water-ice molecules in our solar system \citep{Encrenaz2008,Cleeves2014}. For future space missions, one important goal is to detect signatures of water-ice on or beneath the Moon's surface, as the Moon is Earth’s only natural satellite and its nearest celestial object \citep{Anand2010,Anand2014,Robinson2014}. The discovery of water-ice in the lunar regolith can lead humankind towards sustainable, futuristic space bases in many ways. For example, water-ice can be broken down into Oxygen and Hydrogen molecules; Oxygen can be used by onboard astronauts for breathing, while hydrogen can be used as a fuel for onboard spacecraft \citep{Ross2023,Gallbrecht2024}. 
 
On the surface of the Moon, one of the ideal places to find water-ice signatures is the permanently shadowed regions (PSRs) near its south pole \citep{Feldman2001,Spudis2013,Hayne2015,Deutsch2020}, where temperatures remain below $110\,\mathrm{K}$, making it an ideal place to sustain water-ice molecules \citep{Vasavada1999}. Earlier, India's ISRO\footnote{Indian Space Research Organisation (ISRO), India} successfully launched the \texttt{Chandrayaan-1} mission in 2008\footnote{https://www.issdc.gov.in/chandrayaan1.html}, which mapped roughly 95\% of the lunar surface, providing high-resolution mineralogical and chemical data \citep{Goswami2008}. Its moon mineralogy mapper (M3) instrument found absorption at 2.8 $\mu$m thus found evidence of water ($\mathrm{H_2O}$) and hydroxyl ($\mathrm{OH}$) molecules on the lunar surface \citep{Pieters2009}. NASA's\footnote{National Aeronautics and Space Administration (NASA), USA} Lunar Reconnaissance Orbiter (LRO), with the Diviner instrument, mapped lunar surface temperatures and identified cold traps: $T < 110\,\mathrm{K}$ \citep{Paige2010}. Also, NASA's lunar crater observation and sensing satellite (LCROSS) impacted a crater named \textit{Cabeus}, a permanently shadowed region near the lunar south pole, where they found strong evidence for water vapour \citep{Colaprete2010}.

Recently, ISRO successfully launched \texttt{Chandrayaan-3}, an in-situ lunar exploration mission\footnote{https://www.isro.gov.in/Chandrayaan3.html} on 23rd August 2023 from the Satish Dhawan Space Centre, Sriharikota, with a mission duration of 14 Earth days to analyse the in-situ soil and environment. The \texttt{Chandrayaan-3} spacecraft touches down safely on the lunar south pole, carrying stationary devices named the Vikram lander and the Pragyan rover \citep{Durga2023,Mathavaraj2025,Vijayan2025,Shah2026}. Vikram lander's payloads include the following: radio astronomy of the Moon-bound hypersensitive ionosphere and atmosphere - Langmuir probe (RAMBHA-LP), Chandra's surface thermophysical experiment (ChaSTE), instrument for lunar seismic activity (ILSA) and laser retroreflector array (LRA) \citep{Alam2024,Mathew2025a,Mathew2025b,Onodera2025}. Pragyan rover's payloads include the following: an alpha-particle X-ray spectrometer (APXS) and a laser-induced breakdown spectroscope (LIBS), which help to analyse the elemental and chemical composition of lunar rocks and soil in the south pole region \citep{Sridhar2024a,Sridhar2024b,Vadawale2024,Shetty2025}. Compared with earlier water-ice exploration missions, the purpose of the \texttt{Chandrayaan-3} mission was in-situ elemental analysis of the lunar surface, focusing on mineralogy, morphology, hydration, and gravity anomalies \citep{Rajasekhar2024,Jain2026}. \texttt{Chandrayaan-3} did not directly detect evidence for surface water-ice; rather, its ChaSTE measurements indicate that, unlike in lunar equatorial regions, high latitudes can harbour water-ice, making them technically less challenging sites for future lunar volatile exploration \citep{Durga2025}. Next, ISRO scheduled to launch another lunar mission, \texttt{Chandrayaan-4}, in 2028, with a planned mission duration of 14 Earth days to collect and return soil and rock samples from the lunar surface.

\subsection{Why onboard decisions matters?}
To send and receive commands to onboard instruments landed on the Moon's surface or any other celestial object, we need to operate from Earth itself. It takes about 2.56 seconds for a radio signal to travel from the Earth to the Moon and back \citep{Williams2021}. This may be a short time, but it takes approximately 4 to 22 minutes for a radio signal to travel from the Earth to the Mars in one direction only \citep{Kobs2019}. Given the onboard instruments' limited battery capacity, which depends on the mission's science goals, a longer time delay may affect those goals. Moreover, future missions will generate larger datasets, making automation pipelines essential for efficient analysis and utilisation of onboard resources within time constraints. So an automation framework in which onboard instruments can operate autonomously will be the need of the hour. For example, a rover can efficiently decide \textit{where to drill} by itself only to achieve the optimal science goal. The upcoming \texttt{Chandrayaan-4} mission will be based on a similar in-situ resource utilisation \citep[ISRU;][]{Sanders2005,Sanders2022} framework. In principle, to detect water-ice an onboard system must analyse the following three things and send the corresponding data to the Earth using the onboard device's memory (i.e., data storage and processing capacity) efficiently:
\begin{enumerate}[label=\textbf{(\roman*)}]
\item Stability for drilling a specific location. 
\item The surface temperature of that location.
\item The Hydrogen to Oxygen intensity ratio in that location.
\end{enumerate}
For example, the ChaSTE instrument onboard the \texttt{Chandrayaan-3} Vikram lander measured a peak lunar surface temperature of approximately $340\,\mathrm{K}$ at its south pole to a depth of 10 cm \citep{Mathew2025b}. Together with the LIBS instrument data attached to the Pragyan rover, it did not provide direct evidence of water-ice molecules at the lunar south pole \citep{Sridhar2024b}, but rather hints at indirect evidence for water-ice. Thus, a self-efficient decision-making framework will help identify likely drilling locations for volatiles such as water-ice, enabling efficient use of the instruments' technical resources. Now, to find water-ice, any mission needs to efficiently measure the three things mentioned above within its technical capacity; otherwise, there will be severe waste of technological, human, and time resources. Moreover, the mission will not achieve economic viability or its science goal.

ISRU frameworks will be used in upcoming missions \citep{Garcia2025}, such as NASA's \texttt{VIPER} (Volatiles Investigating Polar Exploration Rover) with Griffin lander, with a planned mission duration of 100 Earth days. This is designed to map the distribution and concentration of water-ice in the permanently shadowed region at the Moon's south pole. The \texttt{VIPER} rover will use onboard neutron and near-infrared (NIR) sensors to locate suitable drill sites without sending commands from Earth \citep{Beyer2025}. The similar frameworks will be used in the upcoming ESA's\footnote{European Space Agency (ESA)} \texttt{PROSPECT} (Package for Resource Observation, In-Situ Prospecting for Exploration, Characterisation \& Testing) mission with a planned mission duration of 10 Earth days \citep{Boazman2024}, and the upcoming Roscosmos's\footnote{National Space Agency of the Russian Federation} \texttt{Luna 27} mission with a planned mission duration of 7 Earth days \citep{Turchinskaya2024}, designed to drill into the Moon's south pole to analyse samples for valuable volatiles such as water-ice.

\subsection{Our idea: \textit{DewTwin-Coin} framework}
In this work, we proposed a pipeline for an autonomous onboard decision making framework named \textit{DewTwin-Coin} in which \textit{Dew} stands for water-ice exploration, \textit{Twin} stands for using two onboard instruments' ChaSTE and LIBS data from the \texttt{Chandrayaan-3} mission, and \textit{Coin} stands for simulating the rover coverage area. The main science goals of this framework are as follows:
\begin{enumerate}[label=\textbf{(\roman*)}]
\item To set up a framework in which physics-based decision rules are tested on real-time \texttt{Chandrayaan-3} data rather than on any simulated data for efficient and sustainable onboard decision making.
\item To set up a framework that implements the \textit{find-S learning} algorithm for planetary science studies.
\item To validate that \textit{zero detection} results for valuable volatiles from the onboard decision-making are also important from a science-goal perspective. This provides robustness to our framework. 
\end{enumerate}

\textit{Find-S learning} is a hypothesis-based supervised learning algorithm that determines whether the most specific hypothesis is consistent with the positive examples or not \citep{Mitchell1997}. The core idea of this algorithm is quite simple: it starts with the most specific hypothesis, which examines only the positive samples (i.e., data samples with positive findings), and then it generalises the hypothesis only as much as necessary to include the subsequent positive samples. We preferred the \textit{find-S learning} algorithm over other more complex algorithms like neural networks because of the following reasons: (i) results based on decision rules are physically interpretable easily, (ii) low computational resources are required, and (iii) results never hallucinate, which means the drilling location will be detected only when all the selection criteria are met.

The rest of the manuscript is structured as follows: Section \ref{sec:2} provides a detailed description of the data sample and the methodology used in our work. In Section \ref{sec:3}, we present our results and discuss their implications. Finally, the conclusions of our work are outlined in Section \ref{sec:4}.

\section{Data and Methodology}
\label{sec:2}

\subsection{Dataset}
\label{sec:2.1}
We have used \texttt{Chandrayaan-3} LIBS and ChaSTE instrument data for our onboard decision-making study. These datasets are publicly available at ISRO PRADAN public data archive\footnote{https://pradan.issdc.gov.in}. The description of each of these datasets is as follows: 
\begin{enumerate}[label=\textbf{(\roman*)}]
\item LIBS data: From the LIBS instrument onboard to the \texttt{Chandrayaan-3} Pragyan rover collected spectra of lunar regolith near the south pole landing site for in-situ elemental and chemical composition studies. From this spectral data, we have collected 3165 raw spectra in the wavelength range $350-950\,\mathrm{nm}$ for our analysis. The mean Hydrogen to Oxygen intensity ratio value for this sample is reported as 0.799.

\item ChaSTE data: From the ChaSTE instrument onboard to the \texttt{Chandrayaan-3} Vikram lander collected thermal data from the lunar regolith. Thermal measurements are important for determining the water-ice stability at the south pole landing site. From this thermal data sample, we have collected 387 raw temperature readings for our analysis. The mean surface temperature value for this sample is reported as $271.4\,\mathrm{K}$.
\end{enumerate}

\subsection{Find-S learning based decision rule}
\label{sec:2.2}
In our proposed \textit{DewTwin-Coin} framework the \textit{find-S learning} is implemented as follows:
\begin{itemize}
\item Hypothesis $\rightarrow$ $H:$ ``This location has water-ice"
\item Finding criterion 1 (thermal constraint) $\rightarrow$ $$C_\mathrm{th}: T < 110\,\mathrm{K}$$
\item Finding criterion 2 (chemical constraint) $\rightarrow$ $$C_\mathrm{ch}: 1.7 < \mathrm{H/O} < 2.3,$$
\end{itemize}
where $T$ is the surface temperature and $\mathrm{H/O}$ is the Hydrogen to Oxygen intensity ratio at a specific location. For each data sample (i.e., the surface temperature and the $\mathrm{H/O}$ intensity ratio at a specific location), the algorithm starts with the hypothesis that $H$ needs to be satisfied very strictly. Now, if the rover, using onboard instruments, sees a suitable drilling location that satisfies both of the above criteria, the hypothesis $H$ is kept, and the rover will drill there; otherwise, the algorithm will reject it, so the rover will move to the next location without drilling the current location. 

The choice of the above two water-ice finding criteria is based on the fact that the presence of water-ice at a site on the lunar surface mainly depends on the surface temperature and the $\mathrm{H/O}$ intensity ratio. The reason why we choose a specific range for these factors is described below:

\begin{enumerate}[label=\textbf{(\roman*)}]
\item Thermal constraint: The sublimation rate of water-ice increases exponentially with temperature due to the strong temperature dependence of the ice vapour pressure \citep{Andreas2007}. At sufficiently low temperatures, water-ice sublimation losses are negligible; above this, it sublimates within hours. $110\,\mathrm{K}$ is commonly used as an upper stability threshold for the surface temperature of lunar cold traps by several studies \citep{Watson1961,Andreas2007,Williams2019}. At $110\,\mathrm{K}$, the predicted sublimation rate for water-ice is of the order of $10\,\mathrm{cm\, Gyr^{-1}}$ only \citep{Williams2019}. This stability threshold may vary depending on the thickness and physical state of the ice. Keeping all this in mind, we define the thermal criterion for finding water-ice on the lunar regolith as:
\begin{equation}
C_\mathrm{th} =
\begin{cases}
1 & \text{if} \; T < 110\,\mathrm{K},\\
0 & \text{otherwise},
\end{cases}
\label{eq:1}
\end{equation}
where `1' signifies \textit{drill here} (i.e., potential water-ice location) and `0' signifies \textit{do not drill here} (i.e., no water-ice location). In \texttt{Chandrayaan-3}, the ChaSTE instrument has been used to record the surface temperature of lunar regolith. 

\item Chemical constraint: Water molecules ($\ce{H2O}$) consist of two Hydrogen atoms for every one Oxygen atom, thus have a stoichiometric atomic abundance ratio of, $$\frac{N_{\mathrm{H}}}{N_{\mathrm{O}}} = 2.$$
Now, \texttt{Chandrayaan-3}'s LIBS instrument does not directly count atoms of a molecule, but rather measures its peak intensity, which is proportional to the corresponding molecular abundance \citep{Cremers2013}. Therefore, the \hi\ line at $656.3,\mathrm{nm}$ and the \oi\ triplet line near $777.4,\mathrm{nm}$ are used as indicators for the Hydrogen and Oxygen atoms, respectively \citep{Vogt2022}. Due to the spectrograph noise (i.e., systematic uncertainty), the possibility of soil mixing, etc., the stoichiometric value cannot always be exactly 2. Thus, for our study, we adopted a tolerance of $\pm 0.3$ around the stoichiometric value 2. This is not a crude limit, but rather an adopted operational criterion that adds flexibility around the expected value. Thus, we define the chemical criterion for finding water-ice on the lunar regolith as:
\begin{equation}
C_\mathrm{ch} =
\begin{cases}
1 & \text{if} \; 1.7 < \dfrac{I_{\hi}(\lambda = 656.3\,\mathrm{nm})}{I_{\oi}(\lambda = 777.4\,\mathrm{nm})}< 2.3, \\
0 & \text{otherwise},
\end{cases}
\label{eq:2}
\end{equation}

So, a location will be a probable site for water-ice availability only if both the above criteria are met. In other words, the final decision rule will be the logical AND between the above two criteria (Eqs. \ref{eq:1} and \ref{eq:2}), i.e.,
\begin{equation}
\mathtt{drill\_decision} =
\begin{cases}
1 & \text{if } \{T,\mathrm{H/O}\} \equiv \{1,1\} \\
  & \quad \in C_{\mathtt{th}} \land C_{\mathtt{ch}}, \\
0 & \text{otherwise}.
\end{cases}
\label{eq:3}
\end{equation}
\end{enumerate}

This is how we have implemented the \textit{find-S learning} algorithm in our study using the LIBS and ChaSTE instruments. It accepts the hypothesis if only all the criteria match.

\subsection{DewTwin-Coin framework}
\label{sec:2.3}
The algorithm of our DewTwin-Coin framework is shown in \ref{alg:1}. This is useful for future lunar missions involving autonomous onboard decision-making (e.g., whether to drill at a site or not). For the \texttt{Chandrayaan-3} ChaSTE and LIBS data, the working mechanism of it is as follows:
\begin{enumerate}[label=\textbf{(\roman*)}]
\item First, the lander's ChaSTE instrument will read the surface temperature of a specific location using its sensor fusion module. Then the rover will evaluate the molecular intensities at $656.3\,\mathrm{nm}$ and $777.4\,\mathrm{nm}$ using the attached LIBS instrument to compute the $\mathrm{H/O}$ intensity ratio at that location.

\item Next, based on the surface temperature and $\mathrm{H/O}$ intensity ratio values, the rover's decision core module will decide whether to drill at that location or not (viz. Eq. \ref{eq:3}). The decision will be in the form of a binary output: either \texttt{drill\_decision} $= 1$ (i.e., \textit{drill here}) or \texttt{drill\_decision} $= 0$ (i.e., \textit{do not drill here}).

\item Finally, the rover's coverage module will simulate its own movement as $x_{t+1} = x_t \pm 1$ with the probability of 0.5 (i.e., a random walk model).
\end{enumerate}
In this way, the rover can efficiently explore the lunar surface to find probable drilling locations with feasible water-ice signatures without using any GPS\footnote{Global Positioning System, a satellite-based navigation network} commands from Earth. Due to the algorithm's simplicity, it can run on any standard CPU, though radiation-hardened CPUs are required for the lunar mission's onboard setup.

\begin{algorithm*}[ht]
\caption{\textit{DewTwin-Coin}: an onboard water-ice detection algorithm}
\label{alg:1}
\begin{algorithmic}[1]
\Require 3165 LIBS spectra $\mathcal{S}=\{S_1(\lambda),S_2(\lambda),\ldots,S_{3165}(\lambda)\}$ (where $\lambda$ denotes the wavelength) and mean surface temperature $T$ 
\vspace{0.3em}
\Ensure Binary decision vector $\mathbf{D}\in\{0,1\}$
\vspace{0.3em}
\State Set thresholds: $T_\ast = 110.0\,\mathrm{K} \;\; \text{and} \;\; r_{\mathrm{low}} (= 1.7) < r < r_{\mathrm{high}} (= 2.3)$ (where $r$ is the $\mathrm{H/O}$ intensity ratio)
\vspace{0.3em}
\For{$i \gets 1$ to $3165$}
\vspace{0.3em}
\State $I_\mathrm{H} \gets \sum_{\lambda=655}^{657} S_i(\lambda)$ \Comment{Integrate H-alpha peak}
\vspace{0.3em}
\State $I_\mathrm{O} \gets \sum_{\lambda=776}^{778} S_i(\lambda)$ \Comment{Integrate O-triplet peak}
\vspace{0.3em}
\State $r_i \gets I_\mathrm{H}/(I_\mathrm{O} + 10^{-6})$
\vspace{0.3em}
\If{$T < T_\ast$ \textbf{and} $r_{\mathrm{low}} < r_i < r_{\mathrm{high}}$}
\vspace{0.3em}
\State $D[i] \gets 1$ \Comment{Potential water-ice site}
\vspace{0.3em}
\Else
\State $D[i] \gets 0$ \Comment{No water-ice site}
\vspace{0.3em}
\EndIf
\EndFor
\State \Return $D$
\end{algorithmic}
\end{algorithm*}

\section{Results and Discussion}
\label{sec:3}
The code for \textit{DewTwin-Coin} framework is written in \texttt{Python 3.11}, and we ran it on an Intel Core i7 8-core CPU with 16 GB of RAM. The pipeline for \textit{DewTwin‑Coin} maintained a stable memory footprint throughout execution. The baseline memory usage was 236.74 MB, and after processing 3165 LIBS spectra and 387 ChaSTE records, it was 237.17 MB. The peak increase was only about 0.43 MB, demonstrating that the algorithm scales linearly and does not require large amounts of memory. This memory efficiency is critical for onboard autonomy, where RAM availability is limited. Our total runtime for the entire dataset is approximately 615 seconds, further confirming that the framework is lightweight and suitable for CPU‑only environments. The results from our analysis are summarised in Table \ref{tab:1}.
\begin{table}[ht]
\centering
\begin{tabular}{lr}
\toprule
                             Metric & Value\\
\midrule
              Mean surface temperature & $271.4\,\mathrm{K}$\\
Minimum $\mathrm{H/O}$ intensity ratio & 0.0\\
Maximum $\mathrm{H/O}$ intensity ratio & 1.786\\
   Mean $\mathrm{H/O}$ intensity ratio & 0.799\\
   Predicted potential water-ice sites & 0\\
                   Predicted dry sites & 3165\\
                Ground truth cold trap & 0\\
                        True negatives & 3165\\
                       False positives & 0\\
         Overlap with ChaSTE cold trap & 0.0\%\\
\bottomrule
\end{tabular}
\caption{Performance of \textit{DewTwin-Coin} on the \texttt{Chandrayaan-3}'s LIBS and ChaSTE data, where positive and negative signify a site is classified as a potential water-ice location or not.}
\label{tab:1}
\end{table}

\begin{figure*}[ht]
\centering
\includegraphics[width=\textwidth]{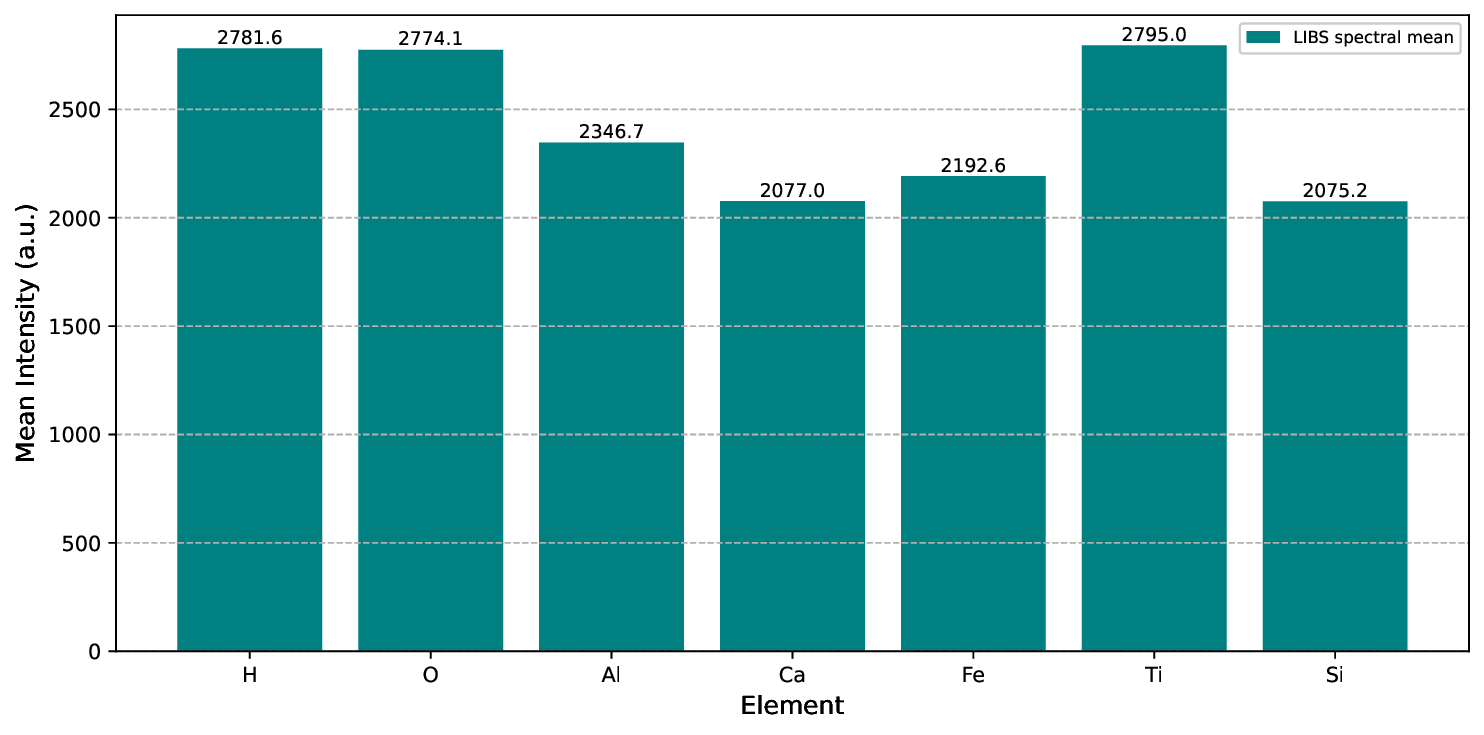}
\caption{Mean elemental intensities measured in arbitrary units \citep[a.u.;][]{Sridhar2024b} from the LIBS data.}
\label{fig:1}
\end{figure*}

In Figure \ref{fig:1} we have shown the mean elemental intensities for elements like Hydrogen ($\mathrm{H}$), Oxygen ($\mathrm{O}$), Aluminium ($\mathrm{Al}$), Calcium ($\mathrm{Ca}$), Iron ($\mathrm{Fe}$), Titanium ($\mathrm{Ti}$), Silicon ($\mathrm{Si}$) measured from the LIBS data. From this, it is evident that in lunar regolith the highest intensities are observed for $\mathrm{H}$, $\mathrm{O}$ and $\mathrm{Ti}$. In contrast, the competitively lower intensities are observed for $\mathrm{Al}$, $\mathrm{Ca}$, $\mathrm{Fe}$ and $\mathrm{Si}$. The presence of a good amount of $\mathrm{Ti}$ indicates that lunar regolith has formed mainly from basaltic rocks. On the other hand, the presence of $\mathrm{H}$ and $\mathrm{O}$ indicate water-ice may exist. For this sample, we found that the minimum and maximum $\mathrm{H/O}$ intensity ratios were 0.0 and 1.786, respectively. But the mean $\mathrm{H/O}$ intensity ratio is low $\sim 0.799$, which should be between 1.7 and 2.3 for water‑ice existence (viz. Eq. \ref{eq:2}). Moreover, the corresponding mean surface temperature from ChaSTE readings is $271.4\,\mathrm{K}$. These all indicate that there might exist some \ce{OH} molecules but not molecular \ce{H2O} \citep{Mccord2011}. In simpler words, the likelihood of water-ice is too low, and most of it may sublimate in the hot terrain. 

\begin{figure*}[ht]
\centering
\includegraphics[width=\textwidth]{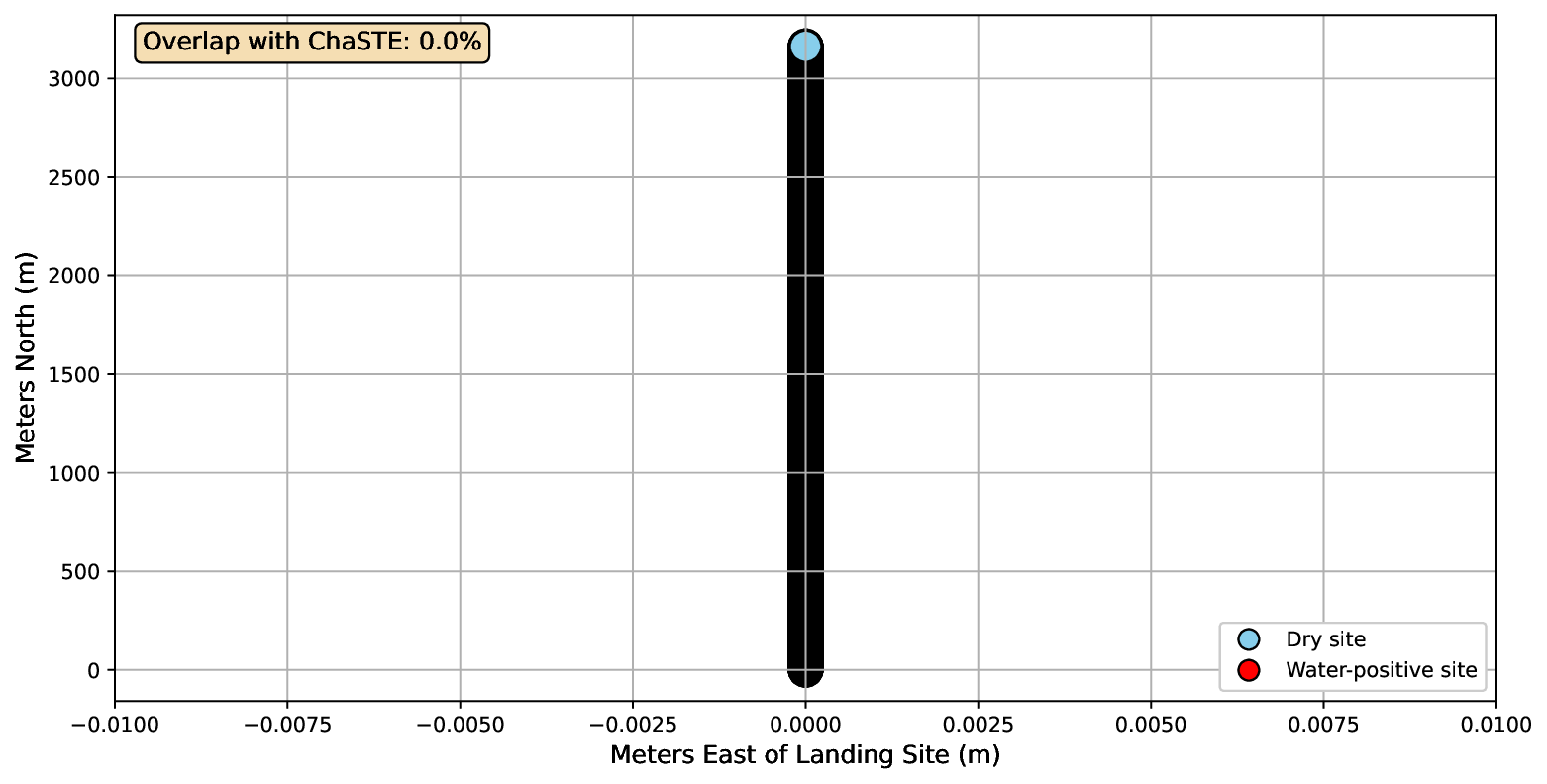}
\caption{\textit{DewTwin-Coin}'s water prospect map, where the $x$ and $y$ axes represent distance east and north of the lunar landing site, respectively. The points marked in blue and red represent dry sites with no water-ice and water‑positive sites, respectively.}
\label{fig:2}
\end{figure*}

In Figure \ref{fig:2}, we have shown the simulated 3165 step path for the rover. In this figure, the vertical alignment of points reflects the rover’s systematic, step-wise movement, inspired by robotic behaviour such as a cleaning robot that moves along a structured path, checking conditions at each step. In simpler words, a robotic vacuum cleaner moves across a floor, marks spots with dirt, and skips clean areas. Similarly, the rover marks sites with water‑ice (if found) and skips dry ones. Now, from the, it is evident that no sites satisfied both the thermal and chemical criteria for water‑ice, i.e., \textit{DewTwin-Coin} yield $\mathtt{drill\_decision} = 0$ (see Eq. \ref{eq:3}) for all the 3165 sites without a single false positive. Thus, results classify all 3165 sites as dry (without water-ice), which is consistent with the \texttt{Chandrayaan-3}'s in-situ readings.

\section{Conclusion}
\label{sec:4}
In this work, we have described an onboard decision-making framework, \textit{DewTwin-Coin}, to assist future lunar missions, such as \texttt{Chandrayaan-4}, in efficiently detecting volatiles, such as water-ice, without relying on GPS commands from Earth. Our framework is based on a simple supervised learning algorithm called \textit{find-S learning} (described in Section \ref{sec:2.3}), which yields decisions using data from the \texttt{Chandrayaan-3} LIBS and ChaSTE instruments. For that, we have collected 3165 LIBS spectra and 387 ChaSTE temperature records with a mean surface temperature of $271.4\,\mathrm{K}$ and a mean $\mathrm{H/O}$ intensity ratio of 0.799. The \textit{DewTwin-Coin} classified all the lunar site locations as dry sites (i.e. water-ice scarcity sites) at the lunar south pole, achieving 100\% agreement with \texttt{Chandrayaan-3}'s in-situ readings regarding the same. This signifies the robustness of our onboard decision-making framework for detecting the locations of volatiles such as water-ice. Also, the results are easily interpretable, and the framework takes about 615 seconds to decide whether to drill at all 3165 locations or not. Thus, it enables efficient use of the mission's technical resources. Overall, the \textit{DewTwin-Coin} can be helpful for future lunar missions in the following ways:
\begin{enumerate}[label=\textbf{(\roman*)}]
\item Instead of waiting roughly 2.56 seconds for receiving GPS command from the Earth, the mission's onboard rover can decide where to drill by itself by running \textit{DewTwin-Coin} on its CPU within a $10\,\mathrm{ms}$ only.

\item Also, using \textit{DewTwin-Coin} means that instead of sending a few megabytes of spectrum data to Earth, it is sufficient to send a 1-bit signal: ``0" or ``1" by the mission's onboard rover that can be received instantly on Earth. Thus, signal bandwidth can be used efficiently over the mission's lifetime.

\item Moreover, our \textit{DewTwin-Coin} framework helps avoid drilling at locations where the likelihood of finding water-ice is negligible. In other words, it helps avoid drilling in locations with hot soil, where water-ice may already have evaporated. Because drilling at a location with temperature and $\mathrm{H/O}$ intensity ratio beyond the thresholds (viz. Eq. \ref{eq:3}) would waste roughly the energy about $500\,\mathrm{W}$ and returns no useful information. Thus, \textit{DewTwin-Coin} is a perfect example of an ISRU framework. 

\item Overall, no water‑positive sites were detected as the \textit{DewTwin-Coin} avoided false positives. Moreover, it demonstrated how a rover can autonomously explore the lunar surface using onboard instruments efficiently both in terms of power and memory capacity. Since this framework is quite simple and effective, it is scalable for future similar lunar missions such as \texttt{Chandrayaan-4}, \texttt{VIPER}, \texttt{PROSPECT}, \texttt{Luna 27}, etc. Additionally, it can be used to find other important volatiles other than water-ice, such as rare earth minerals, for similar planetary missions (e.g., Mars, Venus).
\end{enumerate}

Besides that, our \textit{DewTwin-Coin} framework has some limitations also:
\begin{enumerate}[label=\textbf{(\roman*)}]
\item In this work, we have used the \texttt{Chandrayaan-3}'s LIBS and ChaSTE instruments' data to study the feasibility of finding water-ice at 3165 locations near the \texttt{Chandrayaan-3}'s landing site. Our analysis did not report any site with probable water-ice availability, which is consistent with \texttt{Chandrayaan-3}'s in-situ analysis. \texttt{Chandrayaan-3}'s surface observations were mainly made during the lunar daytime (i.e., the sunlit period) as the mission was designed with solar-powered instruments and the rover depended on sunlight. So, for future lunar missions, once we have PSR site data (which evidently contains water-ice), we can make our \textit{DewTwin-Coin} framework more robust and foolproof by running it on those samples.

\item Also, to determine whether a drilling location contains water-ice or not, we have used fixed thresholds for the surface temperature and the $\mathrm{H/O}$ intensity ratio. But, in principle, there might be some uncertainty from onboard instrument readings \citep{Wiens2012}. Thus, adding uncertainty to the model may help us obtain more robust predictions. This will be a focus area for our future work.

\item Moreover, we have used a basic rover coverage model, i.e., a random walk. But, in principle, rover movement can be restricted by hazardous terrains due to extreme topographical conditions. This will be another focus area for our future work.
\end{enumerate}

\section*{Data and Code Availability}
The data and codes used in the course of the analysis of our \textit{DewTwin-Coin} framework are publicly available at the following GitHub repository - \url{https://github.com/Soundarya-2004/DewTwin-Coin}.

\section*{Conflict of interest}
Both authors declare that they have no conflict of interest.

\section*{Acknowledgments}
Our work uses the ISRO \texttt{Chandrayaan-3} LIBS and ChaSTE instrument data, which are publicly available in the ISRO PRADAN public data archive maintained by the Indian Space Science Data Centre (ISSDC). The author, SR, would like to acknowledge Dr. E. Saravana Kumar (The Oxford College of Engineering, Bommanahalli, India) for his guidance throughout this research. The author DM acknowledges the Council of Scientific \& Industrial Research (CSIR), Government of India, for providing a research grant via an associateship (File No. 09/1178(25506)/2025-EMR-I).
%We also thank the anonymous referee for insightful remarks, which significantly improved the scientific completeness of this manuscript.

%%References section
\bibliography{paper_bib}
\end{document}